\documentclass[11pt]{extarticle}
\usepackage{ftnxtra}
\usepackage{caption}
\usepackage[utf8]{inputenc}
\usepackage{geometry}
\usepackage{amsmath}
\usepackage{float}
\usepackage{graphicx}
\usepackage{dcolumn}
\usepackage{bm}
\usepackage{array}
\usepackage{url}

\begin{document}

\begin{abstract}
In this study, we report a physical model and a Monte Carlo simulation scheme developed to predict the angular distributions of energetic argon atoms and ions as an ion beam passes through a gas-filled volume.
The study explores charge-exchange neutralization as a method for generating fast neutral beams suitable for low-damage, high aspect ratio (HAR) etching. The proposed model and simulation code are straightforward and compact, potentially making them useful tools for prototyping.
\end{abstract}

\title{Ion-neutral and neutral-neutral scattering in argon at KeV energies and implications for high-aspect-ratio etching}

\author{A.~V.~Khrabrov and I.~D.~Kaganovich} 
Princeton Plasma Physics Laboratory, Princeton, NJ 08543

\maketitle
\section{Introduction}
\subsection{Fast atomic beams for high aspect ratio etching}
Fast atomic beams (FABs) have attracted sustained interest in plasma processing and surface engineering as a result of their unique ability to deliver highly directional, energetic neutral particles to a substrate while avoiding many of the deleterious effects associated with the surface charging by charged particle bombardment. Neutral atom beams have been explored for a wide range of materials processing applications, including dry etching, thin-film deposition, implantation, and sputtering. In particular, for plasma-assisted etching, the use of neutral species has long been recognized as an effective means of mitigating surface charge, reducing defect formation, and suppressing undesired inelastic interactions in the gas phase—advantages that were identified in early plasma processing studies and remain central to modern device fabrication today \cite{Coburn1972,Gottscho1992}. 

The continuous development of efficient, well-collimated fast atomic beam sources has, in turn, opened up new opportunities to integrate them into advanced manufacturing systems that require high precision, low defect rates, and reliable performance.
Efficient FAB sources have been realized using a variety of neutralization schemes. At present, many practical implementations rely on surface neutralization on a grid through glancing collisions within narrow slits or channels, including graphite showerheads and capillary-based structures, which can provide high neutralization efficiency and good beam collimation \cite{Ishikawa1998}.
Despite their effectiveness, surface-based neutralizers inevitably introduce surface sputtering, material contamination, and lifetime limitations, issues that become increasingly severe as processing requirements push toward higher FAB energies in the KeV range and extreme aspect-ratio features approaching or exceeding 100:1.
An alternative and increasingly attractive approach is the generation of fast atomic beams through gas-phase neutralization of accelerated ions via charge-exchange collisions. In this process, ions that are accelerated by an electric field within an ion source experience electron-capture collisions with surrounding neutral particles, generating fast neutrals that preserve nearly all of the directed kinetic energy of the original ions. In comparison to surface-based neutralization techniques, gas-phase neutralization provides several benefits, including higher chemical purity, avoidance of sputtered material originating from neutralizer surfaces, and the ability to scale to higher beam energies. The lack of surface interactions is especially important for FABs in the KeV energy range, where surface erosion and contamination can seriously impair beam quality.
We also note that surface neutralizers cannot produce a solid-angle distribution of neutral atoms with a peak on the axis, as a glancing ion-surface collision is necessary to cause an interaction \cite{Cardoso2025}. In contrast, relevant gas-phase differential cross-sections are maximal at the zero angle.
\subsection{Scattering processes in a gas-phase FAB source}
A key performance metric of an atomic beam source utilized in high aspect ratio (HAR) etching is its angular divergence, which is specified by the angular distribution of particle velocities with respect to the beam axis. For advanced processing applications, the divergence must be quite small, with acceptable values typically limited to approximately $1^\circ$ or less \cite{Coburn1972}. Although collimating apertures can be employed to reduce divergence, they do so at the expense of beam intensity and reintroduce surface interactions, thereby undermining the key advantages of gas-phase neutralization. Consequently, minimizing the intrinsic angular divergence of FABs is a central scientific and technological challenge.

\subsection{Experimental work on plasma-based FAB sources} 
Fundamental studies of ion–atom and atom–atom scattering, some of which are cited in subsequent sections, were traditionally performed with accelerated, energy-resolved ion beams extracted from low-pressure (mTorr range or below) hot-cathode glow discharges. For the production of FABs used in etching, ion sources based on high-frequency capacitive or inductive discharges are generally regarded as more suitable. Such sources can easily supply the necessary atomic flux—approximately $1~\mathrm{mA/cm^2}$ when expressed as an equivalent current density \cite{SamukawaReview2022}— and also operate without a consumable cathode. As noted by Dornberg {\em et al.} \cite{Dornberg2023}, this property constitutes a clear advantage over high-power DC sources. In their experimental 
study, those authors further observed a deterioration in etch selectivity with prolonged operation, ultimately attributed to oxides and/or halides of the cathode material depositing on the graphite components of the plasma chamber in a commercial ion-beam etching system.

Samukawa and co-authors have designed, studied, and successfully applied atomic beam sources that couple a surface neutralizer with an inductively coupled discharge, as documented in a series of their publications \cite{Samukawa2001, Samukawa2006_1, Samukawa2007_1,Samukawa2007_2, Kubota2010, Miwa2013,Samukawa2015, Ishihara2022, SamukawaReview2022}. These sources showed good performance at low-to-medium aspect ratio etching of nanometer features. Graphite neutralizer surfaces are more effective in neutralizing negative ions, and plasma chemistry was arranged accordingly. For positive ions, a high rate of neutralization, more than 50\%, was also obtained. The angular divergence of the resulting FAB was estimated, apparently based on simulations, to be around $5^\circ$ which is likely near the minimum achievable for surface-neutralizing arrays.

To accelerate the extracted ions to the specified energy of the fast neutral atoms to be produced, high-frequency plasma sources are combined with an additional DC or low-frequency RF ion extraction stage. In the latter case, the ions are accelerated by the self-bias potential of the low-frequency stage.

Although, in comparison, the gas-phase neutralization technique has not yet been developed to the level of an industrial process prototype, a series of recent experimental works led by authors of Nagoya University \cite{Ichikawa2021, Kim2025_1, Kim2025_2, Kawamura2025} demonstrated that angular divergence well below \(1^\circ\) can be achieved for accelerated ions and for the resulting fast neutral beam, and at sufficiently low pressure only limited by the transverse thermal motion of the beam ions at the source. In these experiments a dual-frequency CCP discharge was employed. The acceleration of ions and the generation of fast neutrals in charge exchange occur in the low-frequency sheath. These results are of special interest to us due to the high resolution, better than \(0.1^\circ\), of the angular distributions of ions and neutral atoms acquired with a microchannel plate detector. The high resolution of the experimental angular distributions that are sharply peaked at low pressure allows a comparison with numerical results both within and outside the thermal spread of the central peak. Such comparison is presented in section \ref{sec:sim_results}. In particular, we interpret the ``tail'' observed outside of the main peak in the measured angular distributions as being due to the finite angular spread in ion-neutral and neutral-neutral elastic collisions. The differential scattering in these collisions is determined by the interaction potential. The same applies to the angular distribution of the fast neutral atoms produced in charge exchange and to the fast neutrals additionally scattered on the background gas prior to reaching the detector. The peak in the angular distributions is produced by particles scattered at small angles, within the range below the thermal spread where the theoretical DSC increases steeply, and, for ion species, also by particles that experience no scattering at all. 

\subsection{The modeling approach}
\label{subsec:scatmodel}
The present work focuses on the development of a simplified ion–neutral and neutral–neutral scattering model tailored for fast ion and  atomic beam simulations in the KeV energy range. The model is designed to retain the essential physics governing angular scattering—namely, the momentum transfer arising from short-range repulsive interactions—while remaining computationally efficient and straightforward to integrate into Monte Carlo collision (MCC) and particle-in-cell Monte Carlo collision (PIC-MCC) frameworks.

The intrinsic source of angular divergence in gas-phase-generated FABs is the scattering in ion–neutral and neutral–neutral collisions within the neutralization volume. Even relatively infrequent collisions can cause pronounced angular deviations at KeV energies as a result of momentum transfer in short-range, repulsive-core interactions. By contrast, collisions with large impact parameters, where long-range attractive forces are dominant, generally produce deflections well below a fraction of $1^\circ$ and therefore can be ignored in the high-aspect-ratio etching application.

Theoretically, the phenomena under study are well understood (e.g. \cite{Hasted1964, McDaniel1964}), that is, it is known how to determine the relevant quantum-mechanical interaction potentials and the respective differential scattering cross-sections (DSCs). The first-principle methods are mathematically exact, but difficult to make computationally efficient when incorporated into particle-based plasma simulations. At high impact energies such as those considered here, the classical treatment of the scattering is valid, but it still requires numerical integration of the orbits to find the scattering angle for the adopted interaction potential and tabulate it with high resolution as a function of both the energy and the impact parameter \cite{Araki2016}.

Presently, we demonstrate that at least for the argon gas, the pairwise interaction in the regime of interest is accurately described by a simple Born-Mayer exponential repulsive potential. The treatment of scattering on this potential is based on an existing highly accurate analytical approximation, obviating the need to numerically integrate the trajectories. The scattering model is implemented in a compact Monte Carlo simulation tool and applied to representative cases of an argon ion beam neutralization through charge exchange. The resulting framework allows for swift assessment of the angular distributions of FNBs, making it particularly appropriate for parametric investigations.

The remainder of this paper is structured as follows. Section \ref{sec:KeV_scattering} presents the physical approach for developing the ion–neutral and neutral–neutral scattering models and summarizes the key assumptions on which they are built, with an approximate analytical description of the scattering process and its numerical implementation detailed in the appendix \ref{appendix_A}. In section \ref{sec:optimal} we consider the optimal length \(L\)of the neutralization cell in units of the CX free path \(\lambda_{\text{cx]}}\). The quantity being optimized is the fraction of the fast neutral flux scattered within a specified small solid angle. We note that this calculation is directly based on the properties of the applicable two-body interaction potential, and ad hoc models available in the literature will not yield a useful answer. A basic computational model of the CX neutralization cell is introduced in section \ref{sec:configuration}. For this configuration, the results of simulating the generation of FAB in argon gas are given in section \ref{sec:sim_results}. Section \ref{sec:summary} summarizes the work and proposes possible avenues for future extensions of the model.

\section{Ion-atom and atom-atom scattering at KeV energies and the choice of the model}
\label{sec:KeV_scattering}
Experimentally measured differential scattering cross-sections (DSCs) of ions in the KeV energy range elastically scattered in a neutral gas, as well as those of fast neutral atoms produced in charge exchange collisions, are strongly anisotropic with a typical half-width below $1^\circ$. Therefore, a model to describe the FAB formation in the HAR etching application needs to accurately account for the predominant small-angle scattering. At the same time, at very small angles (that is, very high impact parameters), the scattering can be ignored if the scattering angle is within the tolerance called for by the technological application. In the present case, the tolerance parameter for the angular divergence of the beam is set by the value of the aspect ratio of the etched features. For the value of 100, it is roughly $0.5^\circ$. In what follows, the tolerance will be set to that angle in the center-of-mass frame of the colliding pair, resulting in the angle of $0.25^\circ$ in the laboratory frame. The physical predictions to be produced by the model are (1) the fraction of particle flux scattered within the specified angle $\theta_*$ and (2) the angular distribution at $\theta>\theta_*$. An additional desirable feature (3) is that the model be simple to implement within a Monte Carlo simulation code. 

In this work, we argue that at the particle energies of interest the deflection by angles $\theta>\theta_*$ occurs entirely due to interactions within the repulsive range of the two-body interaction potential. It is also shown that, at least in the case of argon, the repulsive core or the repulsive wall of the potential can be accurately represented by a simple Born-Mayer exponential dependence $V_{\text{max}}\exp(-br)$  with the parameters introduced by Abrahamson \cite{Abrahamson63, Abrahamson69}. In this work, the values of $V_{\text{max}}$ and $b$ are found by a linear fit of the logarithm of the function tabulated in \cite{Abrahamson63}, although applying the original published values results in no discernible difference. The integral expressing the deflection angle $\theta$ through the collision energy and the impact parameter is not known analytically, but a highly accurate analytical approximation has been found \cite{Heinrich64}. The resulting numerical procedure meets the stated goals of accuracy and simplicity.

Theoretical and experimentally derived two-body potentials feature an exponential repelling wall of the form $V_{0}\exp(-r/a)$ (Born-Mayer) or  $V_{0}\frac{a}{r}\exp(-r/a)$ (Screened Coulomb, Yukawa) and a long-range inverse power attractive contribution, $r^{-4}$ polarization potential for the ion-atom case or $r^{-6}$ dipole-dipole potential for the atom-atom case. The depth of the potential well (the potential minimum) is a small fraction of 1~eV for atom-atom pairs and on the order of 1 eV for the ion-atom case. For the argon gas considered, the respective values are  0.0124~eV and 1.39~eV or much less, depending on the electronic state of the $\mathrm{Ar-Ar^+}$ system. At the same time, the magnitude of the repulsive potential is in the range of several KeV. For center-of-mass (COM) collision energies on the order of hundreds of eV, the presence of an attractive well is unimportant \cite{Myers92, Ruzic93, Maiorov2009} other than at small deflection angles much lower than $1^\circ$. Since for energies in question those angles fall within the acceptable range for the FAB etching application, the task at hand is to investigate the angular distribution of the beam at higher angles and thus to predict the fraction scattered within the acceptable range.

\subsection{Two-body interaction potential for neutral Ar atoms}
\label{subsec:thepotential}
In light of the preceding discussion, a Monte-Carlo scattering model based on the Born-Mayer potential has been adopted to sample the scattering angle. Multiple ab initio and empirical scattering potentials have been evaluated, along with experimental results, with a focus on the repulsive wall region.

It should be noted that for neutral-neutral collisions, the proper accounting of the repulsive interaction has already been recognized as necessary in areas such as combustion and hypersonic flows, given the shallow depth of the potential well and the lack of a physical basis for the $r^{-12}$ repulsive wall presented by the Lennard-Jones potential. In that context, our approach can be categorized as a variation of a variable soft sphere (VSS), presently applied also to the ion-neutral collisions. 

\begin{figure}[H]
    \centering
    \begin{minipage}{0.9\linewidth}
    \includegraphics[scale=0.9]{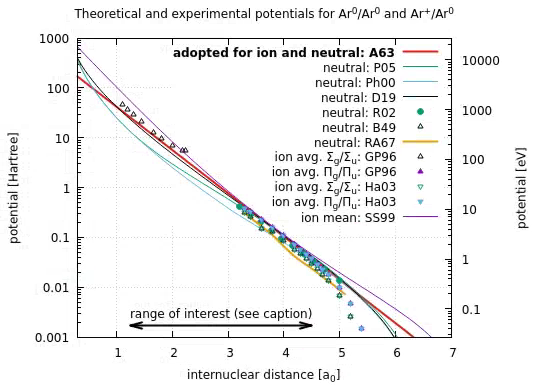}
    \caption{
    A survey of atom-atom and ion-atom interaction potentials for argon at distances below the equilibrium separation of \({\text {Ar}}^0\) pair. The arrowed line spans the range of minimal approach distances relevant to the present problem in the case of 1~KeV projectiles (500~eV center-of-mass collision energy). The lower limit is the head-on approach distance (with \( \theta_{\text{COM}}=\pi\)) and the upper limit is the impact parameter (large enough to be close to the apsis) for which \(\theta_{\text{COM}}=0.5^\circ\). The adopted curve is a Born-Mayer potential with the the parameters based on the work by Abrahamson \cite{Abrahamson63}. Within the uncertainty in the available data, this potential can also be applied to ion scattering on atoms, as discussed in the text. The labels are keyed to the references as A63: \cite{Abrahamson63}, P05: \cite{Patkowsk2005}, Ph00: \cite{Phelps2000}, D19: \cite{SAAP2019}, R02: experiment \cite{PKRol2}, B49: experiment \cite{Berry49}, RA67: experiment \cite{Ross_Adler_1967_2}, GP96: \cite{Gadea1996}, Ha03: \cite{Ha2003}, SS99: \cite{Stefansson99}\footnote{Some numerical entries in \cite{Stefansson99} might be mistyped, given too high values of the potential function at \(r>5 a_0\) where it is expected to be accurate.}
    }
    \label{fig_all_potentials}
    \end{minipage}
\end{figure}

\begin{figure}[H]
    \centering
    \includegraphics[scale=0.9]{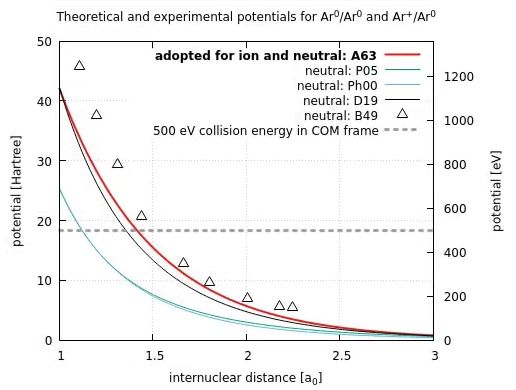}
    \caption{Interaction potentials at closer internuclear distances. The scale is linear and the label keys are as in Fig.~\ref{fig_all_potentials}.
    }
    \label{fig_short_dist_potentials}
\end{figure}

\begin{figure}[H]
    \centering
    \includegraphics[scale=0.9]{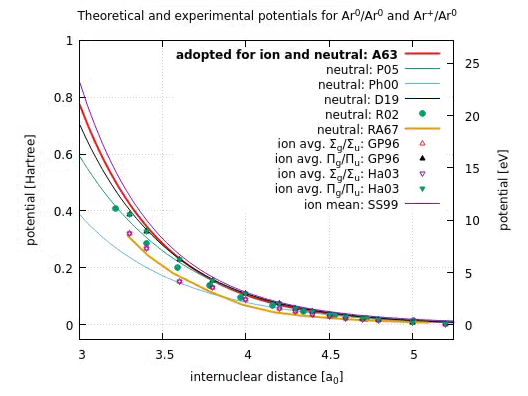}
    \caption{Interaction potentials at intermediate internuclear distances where the repulsive force is still dominant. The scale is linear and the label keys are as in Fig.~\ref{fig_all_potentials}.
    \label{fig_long_dist_potentials}
    }
\end{figure}

A summary of the theoretical models (ab initio and semi-empirical) and experimental results examined for the present study is visualized in Fig.~\ref{fig_all_potentials}.
In Figs.~\ref{fig_short_dist_potentials} and \ref{fig_long_dist_potentials} the two-body potentials in question are displayed, respectively, over shorter and intermediate internuclear distances on a linear scale.

The various continuous curves represent published analytical fits to ab initio and empirical models. The points in the short-distance range show the experimental result of Berry \cite{Berry49} on the Ar-Ar potential. The points in the $r<3a_0$ range map the Ar-Ar experimental data by \cite{PKRol2} and two sets of sate-averaged ion-atom potentials based on the ab-initio theories of Gadea and Paidarová \cite{Gadea1996} and of Ha {\em et al.}\cite{Ha2003}. 

The potential curve selected for use in the scattering model is the Born-Mayer-type analytical fit of Abrahamson \cite{Abrahamson63}. It is found to be in good agreement with a recently derived ab initio SAAP potential \cite{SAAP2019}, and with two sets of experimental data obtained at short and intermediate internuclear distances. 

\subsection{Applying atom–atom potential to ion–atom repulsive interaction}
Two ab initio mean potentials corresponding to the states $\Pi$ and  $\Sigma$ of the ${\text {Ar}}^+-{\text {Ar}}^0$ system are mapped in Fig.~\ref{fig_long_dist_potentials} by two sets of tabulated values sourced from \cite{Gadea1996} and \cite{Ha2003}. All of these values are within the Hartree range of less than 1. The analytical fit to the overall mean potential of Stefánsson and Skullerud \cite{Stefansson99}, provided by these authors, is also shown. The latter potential is an empirical fit aimed at reproducing the swarm data at the values of \(E/N\) corresponding to mean energies that are low compared to the range considered presently; therefore, the observed discrepancy at shorter distances, as seen in Fig.~\ref{fig_short_dist_potentials} is expected. 
Given the uncertainties in the available data, we choose to employ a single potential curve to describe the three pairwise interactions ${\text{Ar}}^0-\text{Ar}^0$, $\text{Ar}^+-\text{Ar}^0(\Sigma)$, and  $\text{Ar}^+-\text{Ar}^0(\Pi)$. Consequently, the 2:1 statistical weighting between the $\Pi$ and $\Sigma$ states becomes irrelevant. We also find that the potential corresponding to the more probable $\Pi$ state aligns more closely with the $\text{Ar}^0-\text{Ar}^0$ potential than the potential for the $\Sigma$ state.
 In addition, it will be assumed that all ion-atom collisions occur at impact parameters within the radius $R_{\text{cx}}$ of the charge exchange process as given by quantum theory and that charge exchange (CX) in such collisions occurs with the probability of $1/2$. As will be further discussed in more detail, for larger impact parameters, $p>R_{\text{cx}}$ (and even below), the resulting deflection angle becomes too small to be of practical relevance. In fact, at 1~KeV energy in the laboratory frame, the scattering angle becomes $0.05^\circ$ (at 1/2 the CM angle) at $p=5.4a_0$, smaller than the CX radius of $8.1a_0$. In principle, the CX interaction at these larger $b$ values can be treated as a 50/50 identity switch, but an accurate model is still required for smaller values. 
 Scattering at extremely small angles is also influenced by the thermal motion of the target atoms \cite{Russek1960}, even before the onset of quantum diffraction effects, unless the target gas is maintained at cryogenic temperatures.
 
 The impact parameter $p$ is drawn uniformly from within the circular region defined by $p \leq R_{\text{cx}}$. The corresponding scattering angle $\theta$ is then evaluated as described in the appendix \ref{appendix_A}, and, with probability of $1/2$, it is replaced by $\pi - \theta$.

\subsection{Additional arguments supporting the validity of the scattering model}
\begin{figure}[H]
    \centering
    \includegraphics[scale=0.75]{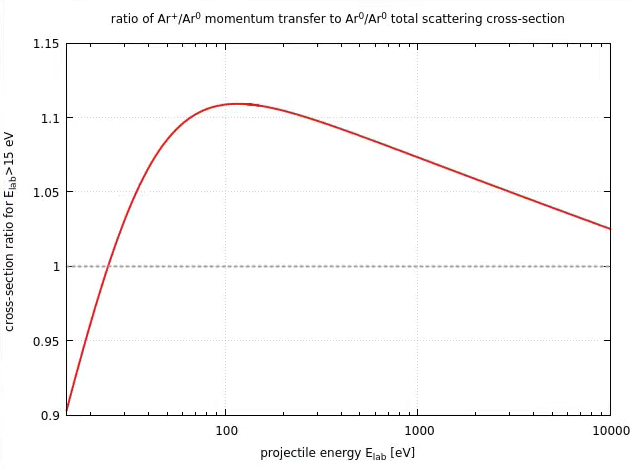}
    \caption{The reference values of the $\text{Ar}^+\text{Ar}^0$ momentum transfer cross-section and of the $\text{Ar}^0\text{Ar}^0$ total cross-section, obtained by a quantum calculation in \cite{Phelps2000}, are quite close at all laboratory frame energies above 15~eV. This observed property is indeed consistent with the underlying assumptions of our model.
    \label{fig_xsec_ratio}
    }
\end{figure}

One consequence of the scattering model used here is that the total ion–atom collision cross section becomes the same as the respective momentum-transfer cross section. For the latter, a reliable value is available. Only collisions at very small scattering angles, where no charge transfer occurs, are omitted, and such angles are irrelevant for the present analysis. 
 Another consequence of adopting identical pairwise potentials for ion-atom and atom-atom scattering is that the respective total cross-sections should be the same because for ions both elastic and CX cross-sections would equate one-half of the atom-atom cross-section. This provides a simple consistency test, which is fulfilled as illustrated in Fig.~\ref{fig_xsec_ratio}.
There, the ratio between the two quantum-mechanical cross-sections is plotted as a function of the projectile energy in the laboratory frame. These integrated cross-sections are accurately known from multiple experimental and theoretical sources and are trusted reference data. The numerical values are from the analytical fits given in \cite{Phelps1994} and \cite{Phelps2000}. The ratio in question is indeed close to unity with a factor of 5\% at energies between 10~eV and 10~KeV. 
In general terms, ion-atom and atom-atom pairwise potentials should be similar in the case of multi-electron atoms interacting within the range where electron screening is the main determining factor.

\begin{figure}[H]
    \centering
    \includegraphics[scale=0.75]{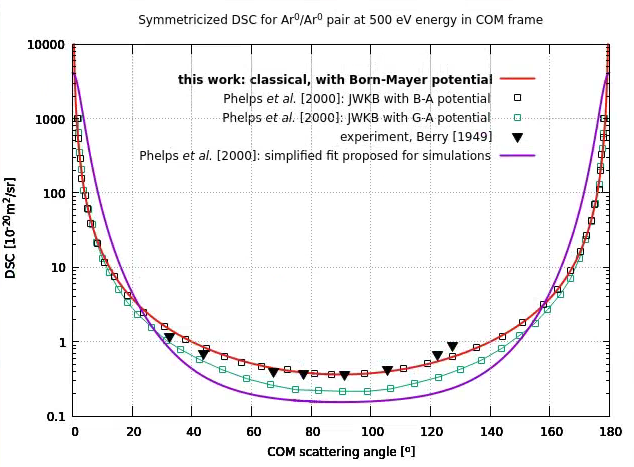}
    \caption{Differential cross-section (DSC) \(\frac{d\sigma}{d\Omega}\) found for the Born-Mayer scattering model is compared to the results of quantum JWKB computations performed by Phelps {\it et al.} \cite{Phelps2000} for two different potentials and to the experimental data of Berry \cite{Berry49} as cited in the former reference. The purple solid line is for the empirical fit recommended in \cite{Phelps2000} for simulation work. The fit is designed to reproduce accurately calculated total, momentum transfer, and viscosity cross-sections. It is seen that the half-width of the resulting DSC is about $1^\circ$ and not applicable to individual collisions. In this context, our result can be viewed as a more accurate quantitative approximation to the underlying quantum calculations of Phelps {\it et al.}.
    \label{fig_Ar0_DSC_verify}
    }
\end{figure}

\begin{figure}[H]
    \centering
    \includegraphics[scale=0.75]{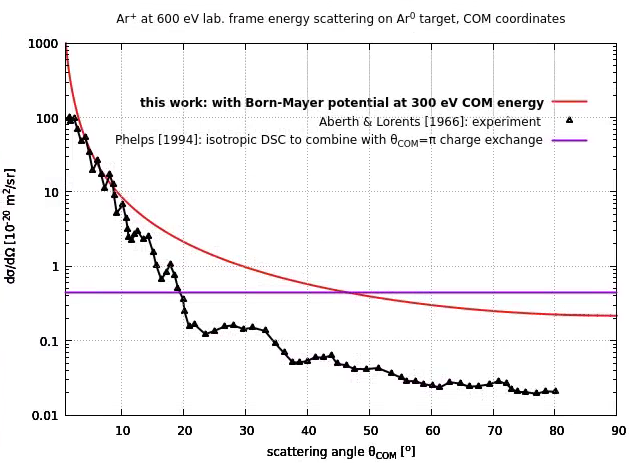}
    \caption {
    Born–Mayer DSC for ions with a laboratory-frame energy of 600~eV, compared with the experimental results of Aberth and Lorents \cite{Aberth1966}. The experimental data have been transformed into the center-of-mass coordinates. Good agreement is found over the key angular range between $1^\circ$ and $10^\circ$. The discrepancy at larger angles remains not fully understood, but may be due to inelastic processes. The horizontal line marks the isotropic scattering cross section recommended by Phelps \cite{Phelps1994} to account for elastic scattering, to be applied in addition to the identity-switch charge-exchange model (with the peak at $180^o$ represented by a $\delta$-function and therefore not shown).
    \label{fig_ArIon_DSC_verify}
    }
\end{figure}

Additional validation of the chosen collision model is established with the help of Figs.~\ref{fig_Ar0_DSC_verify} and \ref{fig_ArIon_DSC_verify}.

\section{Scattering angle vs. impact parameter in the numerical model}
\begin{figure}[H]
    \centering
    \includegraphics[scale=0.75]{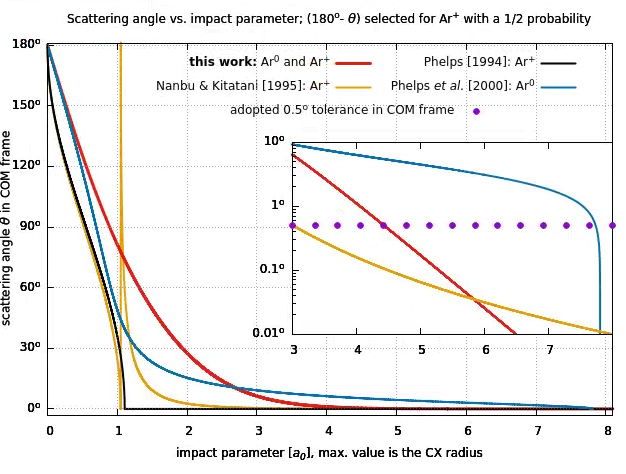}
    \caption{Center-of-mass scattering angle versus impact parameter for the adopted Born–Mayer (BM) potential and two other models. The red curve corresponds to BM elastic scattering of both atoms and ions. Comparison is given against the ion-atom models recommended by Phelps \cite{Phelps1994} and by Nanbu \& Kitatani (NK)\cite{Nanbu95}, and the atom-atom model of Phelps {\it et al.} \cite{Phelps2000}. The elliptic integral $K$ appearing in the NK model was approximated according to \cite{CHACHIYO2026109931}. For the ion particles, charge exchange collision in all cases should be selected by substituting $\theta\rightarrow\pi-\theta$ with a probability of $1/2$.
    \label{fig_Ion_deflection_compare}
    }
\end{figure}

\begin{figure}[H]
    \centering
    \includegraphics[scale=0.75]{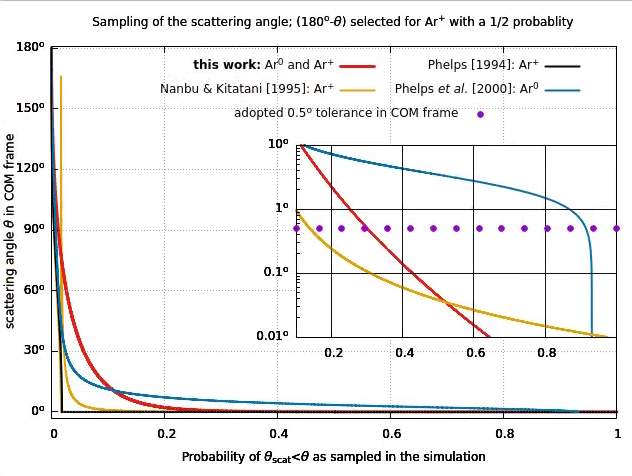}
    \caption {Same functions as plotted in Fig.~\ref{fig_Ion_deflection_compare}, with the impact parameter $p$ replaced by the cumulative probability $(p/p_{max})^2$ of scattering at an angle above $\theta$. The figure therefore illustrates the Monte Carlo sampling of the scattering angle in the COM frame. 
    \label{fig_cumulative}
    }
\end{figure}

For numerical simulations, we calculate the scattering angle for each randomly sampled value of the impact parameter. The underlying calculation is presented in section \ref{appendix_A} in the Appendix. The calculated center-of-mass (COM) scattering angle $\theta$ is plotted in Fig.~\ref{fig_Ion_deflection_compare} versus the impact parameter and in Fig.~\ref{fig_cumulative} versus the corresponding probability of scattering at an angle smaller than $\theta$. The latter plot is a direct illustration of the sampling procedure used in the simulation code. 
In the Phelps \cite{Phelps2000} recommended model of argon ions scattering in the parent gas, the DSC is a sum of two contributions: a delta-function term  representing the identity switch exchange collisions and an additional isotropic term. In turn, the NK model uses scattering based on the induced dipole $r^{-4}$ attractive potential, supplemented with an isotropic term in the regime associated with the non-physical orbiting region. Both the Phelps and the Nanbu-Kitatani models therefore yield an elastic DSC that is strongly peaked at small angles. For very small scattering angles, the deflection predicted by the Nanbu-Kitatani model actually exceeds that of the BM model; however, these angles lie below the angular resolution or the tolerance typically adopted in practical applications, and the associated scattering can thus be neglected.

Fig.~\ref{fig_cumulative} displays the scattering angle as a function of the cumulative probability distribution (i.e. the random number drawn from a uniform distribution on $[0:1]$ used to sample the scattering angle), for a collision energy of 1~KeV in the laboratory frame, or 500~eV in the center-of-mass (COM) frame. 
The model derived based on the BM potential accurately reproduces the structure and width of the backscattering peak at \(180^\circ\), in contrast to the nearly \(\delta\)-function-like peak produced by the models previously recommended. The same applies to the elastic scattering peak at small angles.

Turning to the neutral-neutral collisions, we refer to Fig.~\ref{fig_cumulative} to notice that the model of \cite{Phelps2000} predicts $\theta>1^\circ$ in the COM frame to occur with a probability of about 90\%, underestimating the scattering at smaller angles in the range of interest to the HAR application. This property is also indicated by the corresponding (purple) curve in Fig.~\ref{fig_Ar0_DSC_verify}. The approximately $1^\circ$ screening angle is a fitting parameter resulting from the adopted analytical fit. \footnote{The model of \cite{Phelps2000} might perform better at higher energies, such as in the simulations presented in \cite{wrongCai2025}, although in that work the analytical fit is applied outside of its intended range of validity, while the model of \cite{Phelps1994} is still used for the ions.}

Once the scattering angle dependence on the impact parameter is known, it can be applied not only in numerical simulations, but also to produce a prediction of the optimal length of (or pressure in) the neutralization chamber. This calculation, which is performed analytically with additional simplifying assumptions, is presented next.

\section{Optimal pressure in the gas neutralizer}
\label{sec:optimal}
The length \(L\) of the neutralization chamber, represented as a ratio to the mean free path of charge-exchange \(\lambda\), can be optimized once a quantitative objective has been established that characterizes the desired properties of the fast neutral beam. A natural objective is to maximize the flux of scattered atoms whose deflection angle in the laboratory frame is smaller than a prescribed value $\theta_*$. This angle expresses the acceptable tolerance for the process and, as shown previously, for a 1~KeV primary ion beam, the Born–Mayer scattering model accurately describes the behavior when $2\theta_*>0.5^\circ$, where $2\theta_*$ denotes the associated scattering angle in the center-of-mass frame. 

\begin{figure}[H]
    \centering
    \includegraphics[scale=0.75]{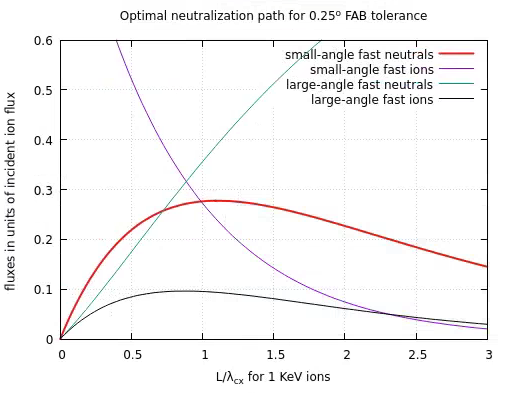}
    \caption{
    \label{fig_optimal_length}
     The particle fluxes as introduced in the text, normalized by the primary ion flux, vs. $L/\lambda_{\text{cx}}$, where $L$ is the length of the neutralization volume and $\lambda_{\text{cx}}$ is the free path with respect to charge exchange collisions. The red curve represents the flux of fast neutral atoms scattered within the acceptable range. 
     }
\end{figure}

To move forward, we define $\sigma_* = \pi b_*^2$ as the cross-section of ion scattering at angles smaller than $\theta_*$ in the COM frame, for the Born-Mayer interaction potential as given in Section \ref{subsec:thepotential}, with $\theta_*=\theta(b_*)$. The cross-section of charge exchange, treated in our model as an independent parameter, is denoted by $\sigma_{\text{cx}}$, and we have $\sigma_*<\sigma_{\text{cx}} = \pi R_{\text{cx}}^2$ because $R_{\text{cx}}$ is several times larger than $b_*$. In the specific case under consideration, of a 1~KeV ion beam energy in the laboratory frame and a chosen value of $\theta_* = 0.25^\circ$, we obtain $R_{\text{cx}} \approx 8.09\,a_0$ and $b_* \approx 4.431\,a_0$. An assumption will be made that upon two consecutive collisions with scattering at $\theta<\theta_*$, the compounded angle of deflection off the beam axis will also be smaller than $\theta_*$. This amounts to neglecting elastic scattering when $b>b_*$ while still taking into account the charge exchange for ion projectiles. The cumulative contribution of repeated small-angle collisions will be neglected, since in the current context the parameter $\alpha = L/\lambda_{\text{cx}}$, defined as the ratio of the neutralization length $L$ to the mean free path of charge-exchange $\lambda_{\text{cx}}$, is of order one. Under these conditions, multiple collisions with $n \geq 3$ are infrequent as they occur with probability $P_n = e^{-\alpha}\sum_{0}^{n}\frac{\alpha^k}{k!}$.
We can now formulate the expressions for the forward fluxes. For the fast neutrals, we additionally observe that the cross-section for scattering within $\theta_*$ is $2\sigma_*$ (with no charge exchange). Define $\alpha = n\sigma_{\text{cx}}$ and $\gamma = n\sigma_* < \alpha$, where $n$ denotes the gas density in the neutralizer. Note that both $\sigma_{\text{cx}}$ and $\sigma_*$ are well defined measurable quantities and that scattering is significant only for $b < b_*$, where the adopted Born–Mayer potential model is still valid. Then
\begin{equation}
    \frac{d\Gamma_i^{\text{fwd}}}{dx}=-(\alpha+\gamma)\Gamma_i^{\text{fwd}},
    \label{eq:ionflux}
 \end{equation}  
\begin{equation}
\frac{d\Gamma_n^{\text{fwd}}}{dx}=(\alpha-\gamma)\Gamma_i^{\text{fwd}}-2\gamma\Gamma_n^{\text{fwd}}.
\label{eq:neutralflux}
\end{equation}
Eq.~(\ref{eq:ionflux}) accounts for fast ions lost in change exchange collisions and for those scattered elastically at angles $\theta>\theta_*$. In turn, Eq.~(\ref{eq:neutralflux}) accounts for the production of forward-going neutrals, identified as those with $\theta<\theta_*$, and for their loss in large angle scattering. The factor of 2 in the r.h.s. appears because, in our model with a single scattering potential and symmetric ion-atom DSC, the atom-atom scattering cross-section equals twice the charge-exchange cross-section (see also Fig.~\ref{fig_xsec_ratio}).
 The ratio of $\Gamma_n^{\text{fwd}}(x)$ evaluated at $x=L$ to the incident primary beam flux $\Gamma_i^{\text{fwd}}(0)$, obtained by solving Eqs.~(\ref{eq:ionflux}) and (\ref{eq:neutralflux}), displays a maximum as a function of the length of neutralization $L$. The optimal value is given by
 \begin{equation}
     L_{\text{opt}}=\frac{1}{\alpha-\gamma}\ln\left(\frac{\alpha+\gamma}{2\gamma}\right)
     \label{eq:Lopt}
 \end{equation}
 and the scaled maximal flux of narrowly scattered fast neutrals is
 \begin{equation}
     \frac{\Gamma_n^{\text{fwd}}(L_{\text{opt}})}{\Gamma_i(0)}=\left(\frac{\alpha+\gamma}{2\gamma}\right)^{-\frac{2\gamma}{\alpha-\gamma}}-\left(\frac{\alpha+\gamma}{2\gamma}\right)^{-\frac{\alpha+\gamma}{\alpha-\gamma}}.
     \label{eq:maxflux}
 \end{equation}
  For the present case, where $\theta_*=0.5^\circ$ implies $\gamma/\alpha=\sigma_*/\sigma_{\text{cx}}\approx 0.3$, the optimal value of $L/\lambda = L\alpha$ given by Eq.~(\ref{eq:Lopt}) is close to unity ($\approx 1.1$), and the corresponding flux ratio calculated from Eq.~(\ref{eq:maxflux}) is $0.28$. Thus, just over one quarter of the initial ion flux can be transformed into fast neutrals that arrive at the target with deflection angles lying within the prescribed tolerance. The main purpose of this calculation is to demonstrate a specific way to quantitatively optimize the product $nL$. 
  However, most of the fast neutrals that contribute to $\Gamma_n^{\text{fwd}}$ will still scatter at very small angles. The solution is illustrated in Fig.~\ref{fig_optimal_length}.
Those fast neutrals not further scattered by the background gas retain the angular distribution imposed, through momentum conservation, by the angular distribution in the charge-exchange event: an ion deflected by an angle $\pi-\theta$ in the center-of-mass frame generates a neutral whose scattering angle is $\theta$, corresponding in the laboratory frame to the incidence angle of $\theta/2$ registered by the detector. Because the DSC for neutral–neutral collisions is identical to that for ion–neutral collisions (apart from the charge-exchange channel), the angular distribution of primary charge-exchange neutrals is therefore the same as that of elastically scattered primary ions.
\section{Simulations of the ion beam neutralization process}
\label{sec:configuration}
\begin{figure}[H]
    \centering
    \includegraphics[scale=0.75]{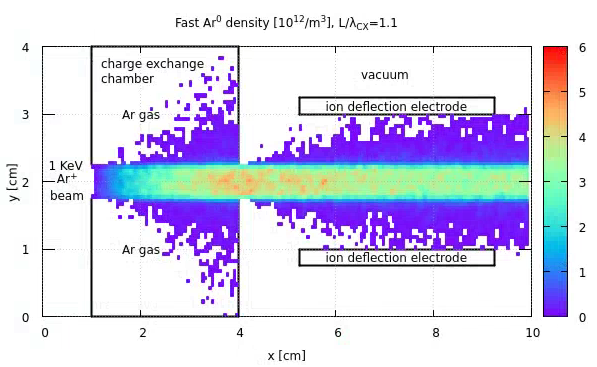}
    \caption{Generic model of a charge-exchange ion beam neutralization cell, showing the neutral beam generated by 1~KeV ions, at the gas density where the travel length equals the optimal value as defined in section \ref{sec:optimal}.
    \label{fig_CX_chamber}
    }
\end{figure}

The mock-up numerical model of the charge-exchange cell used in this study is depicted in Fig.~\ref{fig_CX_chamber}. It is a two-dimensional setup that is used to run simulations with existing code. The space charge was neglected at this time, to focus on collisional transport only. Deflection electrodes are present to direct ions away from the target; however, they are not operated in practice because in the numerical study we also characterize the angular distributions of the ions. The object of interest are the angular distributions of the particles arriving at the right boundary. Strictly speaking, in our study it is sufficient to implement the Monte-Carlo collision model only in the velocity space. The mechanical aperture limits the angle of incidence in the \(xy\) plane to approximately \(10^\circ\). This is quite sufficient, as the angular distributions fall off quite sharply in the low-pressure regime we are interested in. 
A beam of ions at energy \(E_b\) is continuously injected at a constant rate through a 5 millimeter wide segment centered on the left boundary. The simulation proceeds until a steady state is reached and the statistical uncertainty is reduced to a sufficiently low value. The specific numerical values of the simulation parameters are not essential, since the electric field is not computed; the only crucial requirement is that the time step be kept much smaller than the characteristic collision time.
Except for the cold beam example presented for benchmarking in Fig.~\ref{fig_OptCaseResult}, the thermal divergence of the primary beam was set at $T_\perp = 0.044~{\mathrm{eV}}$ to match the value specified in \cite{Kim2025_1} because a comparison with this experimental work is presented in section \ref{sec:Nagoya_comparison}. The background gas temperature \(T_g\) was set to the same value. However, the influence of \(T_g\) is negligible compared to that of $T_\perp$. The reason is that for a small scattering angle $\theta$, the lowest-order correction is proportional to $\theta (T_g/E_b)$. In what follows, the simulation results will be presented and discussed.  

\section{Simulated angular distributions}
\label{sec:sim_results}
All simulated distributions are given as solid-angle probability densities, that is, the particle count is obtained by integrating with a weighting factor of $\sin\theta$. The absolute magnitudes are set by the number of ions injected per time step and by the angular resolution chosen for the output data, and not normalized, except for comparison with an experimental result in Fig.~\ref{fig_KimCompare}.
\subsection{Influence of the aperture}
\begin{figure}[H]
    \centering
    \includegraphics[scale=0.93]{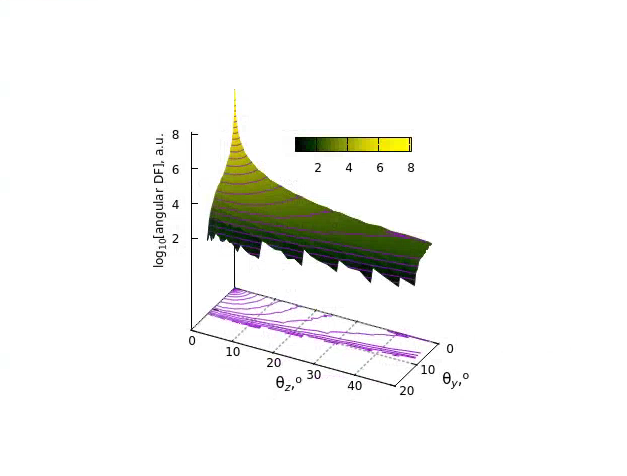}
    \caption{
    \label{fig_2D}
     Two-dimensional distribution of detected fast neutrals over the incident angles $\theta_x$ and $\theta_y$. The angular bin size increases in geometric progression so as to capture the tail, but the first bin is $0.2^\circ\times0.2^\circ$ to provide resolution near the origin similar to that in Fig.~\ref{fig_OptCaseResult}.
        }
\end{figure}

Since the adopted numerical model is two-dimensional in space, it represents a slit aperture with infinite extent in the $z$ direction. Figure~\ref{fig_2D} presents an example of the two-dimensional distribution as a function of the spherical angles $\theta_z$ and $\theta_y$ for fast neutrals, corresponding to the configuration identified as optimal in section \ref{sec:optimal}.
For almost all incoming atoms, the distribution remains azimuthally symmetric, with an anisotropic tail only appearing at angles larger than the $\approx 10^\circ$ value set by the geometrical dimensions shown in Fig.~\ref{fig_CX_chamber}. Such angles are not relevant for the current work. The connection between $\theta_z$, $\theta_y$, and $\theta$ is given by $\sin^2\theta = \sin^2\theta_z+\sin^2\theta_y - \sin^2\theta_z\sin^2\theta_y$. The last term can be neglected because $|\theta_y|<10^\circ$.

\subsection{Angular distributions with cold and warm primary beam}
\label{subsec:coldwarm}
\begin{figure}[H]
    \centering
    \includegraphics[scale=0.8]{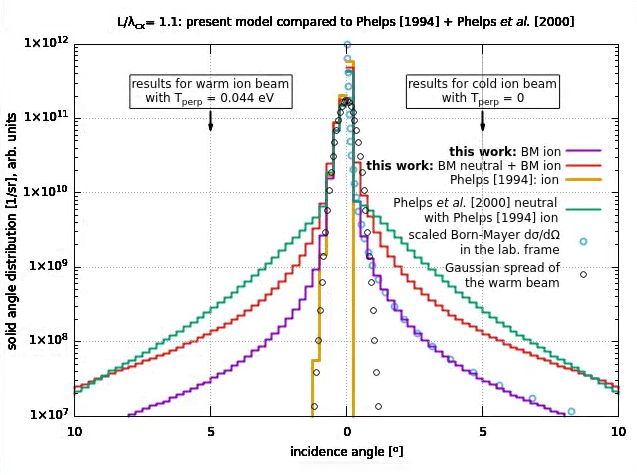}
    \caption{
    \label{fig_OptCaseResult}
    Calculated angular distributions of ions and fast neutrals impacting the right boundary of the simulation domain depicted in Fig.~\ref{fig_CX_chamber}. In addition to the presently developed model, we apply a combination of those recommended in \cite{Phelps1994} for the ions and \cite{Phelps2000} for fast neutrals. The distributions displayed on the right hand side are for the primary ion beam with $T_\perp=0$ and those on the left hand side are for $T_\perp=0.044\,{\text{eV}}$.
    }
\end{figure}

The plots in Fig.~\ref{fig_OptCaseResult} show angular distributions of ions and fast neutral atoms detected at the right boundary of the simulation domain, as seen in Fig.~\ref{fig_CX_chamber}. The distributions are not spatially resolved. In this first example, the density of argon gas within the neutralization region is chosen to maximize the ratio of the flux of narrowly scattered fast neutrals, as defined in section \ref{sec:optimal}, to that of the injected ions. We recall that the optimal condition is \(L/\lambda_{\text{cx}}=1.1\). with single collisions being prevalent.
The angular scale is mirrored with respect to \(\theta = 0\). The right-hand portion of the plot shows the results for an ion beam with \(T_\perp = 0\), used to validate the expected distribution at very small angles. The angular resolution (bin size) in Fig.~\ref{fig_OptCaseResult} is set at \(0.25^\circ\) so the distribution values outside the first bin are physically valid as discussed in section \ref{subsec:scatmodel}. The ion distribution in the cold beam case is compared with the differential scattering cross-section, scaled to obtain a visual fit. The fit is good because single collisions still prevail for \(L/\lambda_{\text{cx}}=1.1\), especially among the particles detected with smaller incident angles.
Alongside the Born–Mayer model used in this work, we also show the distributions derived from combining the \(\text{Ar}^+\) model of Phelps \cite{Phelps1994} with the \(\text{Ar}^0\) scattering model of Phelps {\em et al.} \cite{Phelps2000}, both of which are discussed in section \ref{subsec:scatmodel}. The Nanbu–Kitatani ion scattering model is excluded from the simulations because, like the Phelps model (see Fig.~\ref{fig_cumulative}), it fails to predict appreciable elastic scattering of ions and thus, through the conservation of momentum, also the angular distribution of fast neutral atoms originating in charge exchange.

The distributions obtained with the model introduced in this work are plotted using purple for ions and red for fast neutral atoms. The distributions based on \cite{Phelps1994, Phelps2000} are plotted with orange for ions and green for neutrals. The observations to be noted are as follows. 
\begin{enumerate}
\item In elastic scattering, the distributions of both ions and fast neutrals predicted by our model are, by design, of the same shape. At low pressure, it is given by the differential cross-section presented in the center-of-mass frame in Fig.~\ref{fig_Ar0_DSC_verify} and transformed to the laboratory frame for Fig.~\ref{fig_OptCaseResult}. The distributions of both ions and fast neutrals possess a finite measurable scale. 
\item For the model recommended in \cite{Phelps1994} the number of elastically scattered ions is too small to be seen in the distributions within the range chosen for the plot in Fig.~\ref{fig_OptCaseResult}. The ions which have not undergone charge exchange collisions are un-scattered and the apparent width of their distribution is set either by the angular resolution of the detector (cold case in Fig.~\ref{fig_OptCaseResult}) or the thermal spread of the beam (warm case in the same figure), whichever is larger. 
\item On the other hand, the distribution of \(\text{Ar}^0\) atoms predicted by the model recommended in \cite{Phelps2000} and used in conjunction with the ion model of \cite{Phelps1994} is wider than that based on the Born-Mayer potential, with a half-width of about \(1^\circ\). It is a consequence of employing screened-Coulomb DSC under Born approximation, in combination with isotropic one, to fit accurately calculated total, momentum transfer, and viscosity cross-sections. The underlying reason is that all three cross-sections diverge for the unscreened Coulomb scattering and the fitted screening angle becomes artificially high. This topic will not be addressed here in further detail. The number of particles scattered with \(1^\circ<\theta<5^\circ\) is approximately twice the number for the Born-Mayer model, with a correspondingly lower number at \(\theta<1^\circ\). For the low pressure case being examined and a cold ion beam (right half of the plot), the two models produce close values of the angular distribution for narrowly scattered atoms \(\theta<0.25^\circ\). The reason is that at 1~KeV the ion scattering model of \cite{Phelps1994} is essentially an identity-switch charge exchange, as mentioned above, and those fast neutrals not scattered further retain the \(\delta\) function angular distribution of the ion beam. At higher pressure, with multiple collisions taking place, the respective predicted distributions of \(\text{Ar}^0\) become markedly different, as demonstrated below in section \ref{subsec:pressure_dependence}.
\item The distributions plotted in the left half of Fig.~\ref{fig_OptCaseResult} show the influence of the thermal spread of the perpendicular velocities of the beam ions. An important observation is that the accuracy  of the scattering model at very small angles becomes less important. The differential cross-section only needs to be accurately known down to the angle on the order of the thermal spread \(\theta_{th}=\sqrt{\frac{T_\perp}{E_b}}\,{\text{rad}}\) or approximately \(0.4^\circ\) in the case at hand. Note that for $T_\perp> 300\,{\text{K}}$ the scattering model based on the Born-Mayer repulsive potential is valid at $\theta>\theta_{th}$ and therefore the predicted distributions with a Maxwellian central peak are physically valid overall.
\end{enumerate}

\subsection{Pressure dependence in our model and for \cite{Phelps1994, Phelps2000}}
\begin{figure}[H]
    \centering
    \includegraphics[scale=0.95]{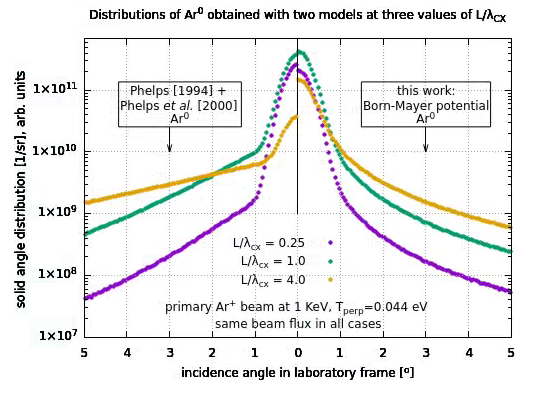}
    \caption{
    \label{fig_NeutralPhelpsCompare}
    make new caption
    }
\end{figure}

To study how the angular distributions of neutral atoms hitting the target vary with the gas density in the neutralizer, we consider three simulations performed for the values of $L/\lambda_{\text{cx}}$ set to 0.25, 1, and 4. The results are presented in Fig.~\ref{fig_NeutralPhelpsCompare} under the same scattering models as in section \ref{subsec:coldwarm}, but this time only for $\text{Ar}^0$ and not for $\text{Ar}^+$. This is because, as has been shown, at high energy ion distributions based on the model of \cite{Phelps1994} do not exhibit a structure but are exponentially decreasing $\delta$ functions with the measured width set by thermal spread or instrumental resolution. 
\label{subsec:pressure_dependence}
The angular scale is again mirrored with respect to the axial angle $\theta=0$. The distributions predicted by our model are plotted on the right, and the comparison case is shown on the left. The main observation is that the Monte Carlo model recommended for $\text{Ar}^0$ in \cite{Phelps2000} predicts a much stronger increase in isotropization in response to increasing pressure. This effect is due to the approximately $1^\circ$ half-width of the respective differential cross-section, already mentioned and visualized in Figs.~\ref{fig_Ion_deflection_compare} and \ref{fig_cumulative}. The said property is especially noticeable in the highest pressure case $L/\lambda_{\text{cx}}=4$, where the number of fast neutrals reaching the target without scattering is small compared to those that did collide.

The model-predicted distributions of energetic argon ions and neutrals can, in principle, be employed for quantitative comparison with experimental measurements. The cases presented in Figs.~\ref{fig_Ar0_DSC_verify} and \ref{fig_ArIon_DSC_verify} constitute representative examples in the low-pressure regime, in which single-collision processes are strongly dominant, given that the simulated distributions have been validated to reproduce the calculated differential cross section, as demonstrated in Fig.~\ref{fig_OptCaseResult}. In Section~\ref{sec:Nagoya_comparison}, we further utilize the present results to support the interpretation of published high-resolution experimental data obtained using a plasma-based RF ion source such as expected to be found in industrial applications. This source is not monoenergetic and there are other challenges to be considered, but a baseline comparison should be worthwhile.   

\section{Comparison with experimental data obtained using RF ion source} 
\label{sec:Nagoya_comparison}
\subsection{Measured distribution vs simulated}
Gas-phase neutralizers for the generation of fast atom beams (FABs) in materials processing are expected to rely predominantly on ion beams extracted from discharge plasmas, for example those produced in dual-frequency capacitively coupled plasma (CCP/CCP) or inductively/capacitively coupled plasma (ICP/CCP) configurations. Such plasma sources, which are already in use in conjunction with surface neutralizers, can deliver wide-area beams with adequately high flux densities.

High-resolution measurements of the angular distributions of ions and fast neutral atoms extracted from a dual-frequency CCP discharge have recently been conducted in a series of experiments at Nagoya University \cite{Ichikawa2021, Kim2025_1, Kim2025_2, Kawamura2025}. In these studies, the ion and neutral beams were extracted from the discharge region through a sub-Debye orifice in the electrode plate and then propagated through a high-vacuum drift tube toward a diagnostic system based on a microchannel-plate (MCP) detector. A rigorous quantitative comparison is difficult to carry out because it would require detailed knowledge of the discharge structure in order to obtain the velocity distributions of ions and fast neutrals. However, a qualitative comparison can still be made based on a key experimental finding: the observed angular distribution is broader than what would be expected from thermal effects alone. In our model, this angular spread is determined by the underlying pairwise interaction potential, as already demonstrated in the panels on the right-hand side of Figs.~\ref{fig_OptCaseResult} and \ref{fig_NeutralPhelpsCompare}.

\begin{figure}[H]
    \centering
    \includegraphics[scale=0.9]{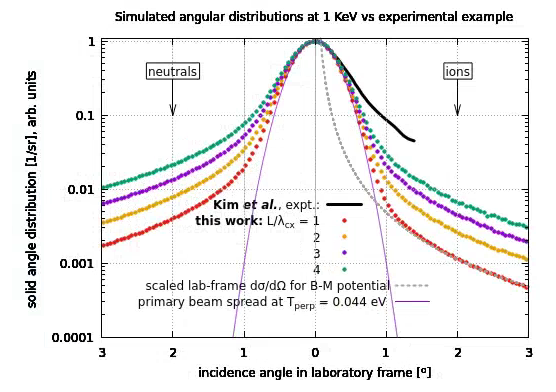}
    \caption{
    \label{fig_KimCompare}
     Solid-angle distributions of ions and fast neutrals impacting the target, in the same format as the experimental data in []. The small-angle peaks are fitted with Gaussian distributions, while the ``wide'' range of the distributions is seen to be due to the specific angular dependence of the differential cross-section for the Born-Mayer interaction potential.
        }
\end{figure}

The particular experimental data set selected for comparison is the ion angular distribution shown in Fig.~4 of \cite{Kim2025_2}. This plot has been digitized and reproduced on the right side of our Fig.~\ref{fig_KimCompare}, where it is displayed together with our simulated angular distributions for four distinct values of $L/\lambda_{\text{cx}}$. In the original analysis, the experimental distributions were effectively characterized by a bi-Gaussian fit, in which the main, narrow peak was attributed to the thermal spread, while a weaker, broader component was interpreted as a ``tail" of still unknown origin. For ions, the authors further adapted this procedure to determine energy-resolved distributions \cite{Kim2025_1}. This was achieved by deflecting the ions and employing a time-of-flight method combined with image analysis of the two-dimensional MCP detector images. In the present work, we are unable to use such energy-resolved distributions since they were not explicitly reported in the cited publications; only the parameters of the bi-Gaussian fits to the solid-angle distributions were provided.
The experimental data in the case at hand were obtained by measuring the flux distribution over a straight line passing through the beam center and converting it into an angular distribution. The angle range was given as $\pm 1.4^\circ$ from the beam axis.

The referenced plot from \cite{Kim2025_2} is given there for illustration,  without citing either the pressure or the width of the accelerating RF sheath (the latter is also the characteristic scale on which the ions are neutralized). However, for $T_\perp=0.045\,\text{eV}$ reported by the authors as the value consistently seen for $\text{Ar}^+$ ions in all cases, the Gaussian curve showing the thermal peak in Fig.~4 of \cite{Kim2025_2} indicates an ion energy close to 1~KeV. Since the measured distribution was normalized by its maximum value at $\theta=0$, our simulated distributions in Fig.~\ref{fig_KimCompare} are normalized in the same manner. The angular scale is again mirrored with respect to $\theta=0$. This time, the ion distributions, including the experimental one, are the ones plotted on the right hand side while the fast neutral distributions are plotted on the left. 
We observe that even at the gas density corresponding to $L/\lambda_{\text{cx}}=4$, likely outside the range explored in the experiment, the simulated distribution still shows a steeper fall-off than the experimentally measured one reported in \cite{Kim2025_2}. Although quantitative agreement is not seen, it would be reasonable to claim that the non-thermal ``tail" observed in the experimental distribution is determined by the applicable pairwise interaction potential. In \cite{Ichikawa2021}, the authors demonstrated that the ``tail" in the ion distribution cannot be accounted for within the scattering model of \cite{Phelps1994}. Such is indeed the case, as the elastic scattering predicted by that model at high energies is much smaller than predicted by the model based on a realistic potential. The differential scattering cross-section for the Born-Mayer model, scaled to visually fit the ion distribution in the lowest pressure case, is traced with a dashed gray line, and the thermal distribution of un-scattered ions is also shown.
For reference, normalized angular distributions of fast neutral atoms are also plotted for each of the four simulation cases. They are seen to be wider than the respective ion distributions, more so at higher gas pressure. We did not attempt to compare these distributions with the measured ones reported in Fig.~6 of \cite{Kawamura2025} at several different values of power deposited in the discharge (the figure is not addressed in the main text), although the comparison would likely have been more favorable than in the ion case.
\subsection{Further discussion}
An important feature of the experimental method using a microchannel plate (MCP) detector is the angular sensitivity of the MCP system. Although this characteristic must have been considered in the design of the experimental setup, the corresponding analysis is not presented in \cite{Ichikawa2021, Kim2025_1, Kim2025_2, Kawamura2025}. Angular-dependent sensitivity may play a more significant role when probing distributions in the broader ``tail'' region than within the relatively narrow range of the thermal spread. An example of an experimental investigation of the angular sensitivity of a microchannel plate is provided in \cite{Gao84_angular_resolution}. In future experiments specifically designed to obtain angular distributions over a broad angular range, it may be beneficial to calibrate the detection system using a particle beam whose angular distribution is independently known and/or to study how the measured distributions vary with respect to the tilt angle of the microchannel plate relative to the beam axis, provided that this angle can be adjusted.

\section{Summary and future work}
\label{sec:summary}
A physical model and a simulation scheme have been developed to predict and analyze the angular distribution of energetic argon neutral atoms produced by charge-exchange neutralization of an ion beam. This model is also applied to ion scattering in elastic collisions and, therefore, due to conservation of momentum, to the angular distribution of fast neutrals generated in charge exchange collisions.
An important takeaway is that the scattering of both ion and neutral species is anisotropic at all energies down to just above the thermal range and needs to be treated carefully in order to obtain meaningful results. The Monte Carlo sampling of the scattering angle is performed through a simple and accurate analytic approximation without a need to numerically integrate the collision orbit.

In the end, future results should be put to use in the design of the actual beam neutralization device. In terms of the model, it will require accounting for the space charge and for realistic ion beam optics. Inelastic collisions should also be taken into account because ion-impact and neutral-impact ionization of the background gas will produce a decelerated tail in the beam distribution measured at the target. 
\section*{Acknowledgments}
This research was supported in part by Samsung and the U.S. Department of Energy through the PPPL CRADA agreement with Applied Materials, the Princeton Collaborative Research Facility (PCRF) under Contract No. DEAC02-09CH11466.
\appendix 
\section{Monte Carlo treatment of scattering in the Born–Mayer potential}
\label{appendix_A} 

For simulation work, a significant advantage to utilizing the Born-Mayer interaction potential is that the scattering angle can be computed analytically with good accuracy, thus eliminating the need for integrating the orbit. The Monte Carlo treatment of collisions in our numerical model is based on an approximate analytical relationship between the apsis (the distance of the closest approach) and the center-of-mass scattering angle $\theta$. This result is due to Heinrich \cite{Heinrich64}, who developed a highly accurate approximation for the scattering angle applicable to a certain class of  repulsive interaction potentials. 

We adopt the following notation: $E_{coll}$ denotes the initial energy of the colliding pair in the center-of-mass frame (which corresponds to one-half of the projectile energy in the laboratory frame for colliding particles of identical mass, neglecting the thermal motion); $V(r)=V_0\exp(-r/a)$ represents the Born–Mayer two-body potential; $b$ is the impact parameter; and $r_{min}$ is the apsis. The corresponding normalized variables are then defined as
\begin{equation}
    \epsilon=\frac{E_{coll}}{V_0}\ll1,
    \label{eq:epsilon_define}
\end{equation}
\begin{equation}
    \beta=\frac{b}{a},
\end{equation}
\begin{equation}
    \rho_{min}=\frac{r_{min}}{a}.
\end{equation}
The above three quantities satisfy the relation
\begin{equation}
    \beta(\rho_{min},\epsilon)=\rho_{min}\sqrt{1-\frac{1}{\epsilon}\exp(-\rho_{min})}
    \label{eq:apsis}
\end{equation}
which implicitly defines the turning point $\rho_{min}(\beta,\epsilon)$ for the Born-Mayer potential field.
For $\rho_{min}\gg 1$ the deflection angle is approximated according to \cite{Heinrich64} as
\begin{equation}
    \theta=\vartheta_1(\rho_{min},\epsilon)=\frac{1}{\epsilon}\frac{\rho_{min}^3}{\beta^2(\rho_{min},\epsilon)}K_0(\rho_{min}),
    \label{eq:theta1}
\end{equation}
where $K_0(\cdot)$ is the modified Bessel function of the second order. In addition, we utilize a compact and accurate approximation of $K_0$ due to Martin and Maass \cite{Martin2022}.
It is evident from Eq.~(\ref{eq:theta1}) that one recovers the large-impact-parameter (momentum) approximation when the scaled impact parameter $\beta$ is used instead of $\rho_{min}$. However, it provides a higher accuracy at moderate angles than the plain momentum approximation, since the dominant contribution to the scattering angle integral arises in the vicinity of $r = r_{min}$. 
With the aid of Eq.~(\ref{eq:theta1}), Heinrich's approximate expression for the deflection angle at an arbitrary impact parameter can be written as
\begin{equation}
    \theta(\rho_{min},\varepsilon)\approx 2\arcsin{\frac{1}{1+2/\vartheta_1(\rho_{min},\epsilon)}}.
    \label{eq:deflect}
\end{equation}
Eq.~(\ref{eq:deflect}) arises from an interpolation between two separate series expansions: one valid in the vicinity of $\theta=0$ and the other valid near $\theta=\pi$. For a projectile energy of 1~KeV in the laboratory frame, we found that the relative error, vs. direct integration over the trajectory, remains below 0.1\% throughout the range $0<\theta\le\pi$. The scaled impact parameter $\beta$ and the deflection angle $\theta$ are parametrically defined by Eqs.~(\ref{eq:apsis}) and (\ref{eq:deflect}) as functions of $\rho_{min}$ and $\epsilon$.
The formal condition of validity seen in Eq.~(\ref{eq:epsilon_define}) is well satisfied for a 1~KeV primary ion beam (500~eV CM frame energy), with $\epsilon\approx 0.06$.
We also define the scaled head-on approach distance $\rho_*$, the apsis for $\beta=0$, as
\begin{equation}
    \rho_*(\epsilon)=\frac{r_*(E_{coll})}{a}=-\ln\frac{E_{coll}}{V_0}=-\ln\epsilon.
    \label{eq:headon}
\end{equation}
It is seen that Eq.~(\ref{eq:deflect}) correctly yields $\theta(\rho_*,\epsilon) = \pi$ in the limiting case of a head-on collision with $\beta = 0$.
To apply Eq.~(\ref{eq:deflect}) to a scattering event in a Monte-Carlo simulation, one first samples the scaled impact parameter $\beta$ according to $\beta=\beta_{max}\sqrt{R}$, where $\beta_{max}=b_{max}/a$. The maximum impact parameter, $b_{max}$, is selected so that the resulting deflection angle is much smaller than the tolerance for the application and/or the angular resolution of the experiment. In the case of ion-neutral scattering, $b_{max}$ is taken to be the charge-exchange radius, as explained in the main text. For neutral-neutral scattering, the most straightforward method is to choose $\pi b_{max}^2$ equal to the known quantum-mechanical cross section. Although the scattering angles for high values of the impact parameter will not be accurately reproduced, they will be sufficiently small to be ignored anyway.
In the next step, Eq.~(\ref{eq:apsis}) has to be solved for $\rho_{min}=\rho_{min}(\beta,\epsilon)$. This is carried out in the same manner as when determining the turning point to integrate the collision trajectory. A numerically convenient version of Eq.~(\ref{eq:apsis}) suitable for applying Newton's method is given by
\begin{equation}
    1-\exp[h(1-\tilde\rho_{min})] - \frac{\tilde\beta^2}{\tilde\rho_{min}^2}=0,
\end{equation}
where \(h=-\log\epsilon\) is the scaled head-on approach distance and \(\tilde\rho_{min}=\rho_{min}/h\ge 1\), \(\tilde\beta=\beta/h\).
The axial scattering angle in the COM frame is then calculated from Eq.~(\ref{eq:deflect}), and in the case of ion–neutral collisions it is replaced by $\pi-\theta$ with a probability of one-half. The azimuthal rotation angle $0\le\phi<2\pi$ is chosen at random and the projectile’s velocity vector is updated using a standard Euler rotation. In the COM frame, the target’s post-collision velocity vector is directed opposite to that of the projectile, and any resulting fast neutral (according to the set energy threshold) is then tracked in the simulation. Finally, the velocities of both collision partners are transformed back to the laboratory frame. With these steps, the scattering procedure is fully defined in the simulation.

For both ion-neutral and neutral-neutral scattering events, the ions and fast neutrals emerging in the collision are tracked as long as their energies remain above a prescribed threshold. This condition prevents an accumulation of slow particles that cannot be modeled reliably without accounting for space-charge effects and gas pumping. In this work, the threshold is chosen as 400~eV, a regime in which the scattering described by the Born–Mayer potential continues to be a valid approximation

\section{Differential scattering cross-section}
\label{Appendix_B}
The angular DSC, parameterized by $\rho_{min}$ and $\epsilon$, is calculated according to 
\begin{equation}
    \frac{d\sigma}{d\Omega}=\frac{b(\rho_{min},\epsilon)}{\sin\theta(\rho_{min},\epsilon)}\frac{\partial b/\partial \rho_{min}}{\partial\theta/\partial\rho_{min}}.
    \label{eq:dsc}
\end{equation}
The DSCs for momentum transfer and viscosity are obtained by multiplying Eq.~(\ref{eq:dsc}) by the corresponding weighting factors. For the Born-Mayer potential, the integrals defining both these transport cross-sections converge. Therefore, their values can be found classically and compared to those presented in \cite{Phelps2000}. Good agreement is observed, after accounting for the fact that the potential adopted in that work is smaller than our adopted Born-Mayer potential at closer internuclear separations $r<2a_0$.
Equation~(\ref{eq:dsc}) was the one used to compute $\frac{d\sigma}{d\Omega}$, which was then tabulated as a function of $\theta$ using Eq.~(\ref{eq:deflect}), producing Figs.~\ref{fig_Ar0_DSC_verify} and \ref{fig_ArIon_DSC_verify}.


\bibliographystyle{unsrt}

\bibliography{references}

\end{document}